\documentclass{article}

\usepackage{cite}
\usepackage{graphicx}
\usepackage{dcolumn}

\begin{document}

\title{On two different kinds of resonances in one-dimensional quantum-mechanical
models}
\author{Francisco M. Fern\'{a}ndez\thanks{%
E-mail: fernande@quimica.unlp.edu.ar}\, and Javier Garcia \\
INIFTA (CONICET, UNLP), Divisi\'on Qu\'imica Te\'orica\\
Blvd. 113 S/N, Sucursal 4, Casilla de Correo 16,\\
1900 La Plata, Argentina}
\maketitle

\begin{abstract}
We apply the Riccati-Pad\'{e} method and the Rayleigh-Ritz method with
complex rotation to the study of the resonances of a one-dimensional well
with two barriers. The model exhibits two different kinds of resonances and
we calculate them by means of both approaches. While the Rayleigh-Ritz
method reveals each set at a particular interval of rotation angles the
Riccati Pad\'{e} method yields both of them as roots of the same Hankel
determinants.
\end{abstract}

\newpage

\section{Introduction}

Several years ago Moyseyev et al\cite{MCW78} discussed the application of
complex rotation to the calculation of resonances. As a simple, nontrivial
illustrative example they chose the potential $V(x)=\left( \frac{1}{2}%
x^{2}-J\right) \mathrm{\exp }\left( -\lambda x^{2}\right) +J$ that exhibits
``pre-dissociating resonances analogous to those found in diatomic
molecules''. The same model was chosen by other authors to test different
approaches for the calculation of resonances\cite{ALR80,REB81,REB82,BEM00}
and a controversy about the behaviour of $\Re E$ vs. $\Im E$ arose\cite
{KLM82,REB82b}. The discrepancy between the results of Rittby et al\cite
{REB81,REB82} and Korsch et al\cite{KLM82} was shown to be caused by the
choice of the rotation angle $\theta $ with respect to the critical angle $%
\theta _{crit}$\cite{REB82b}. The set of resonances that one obtains with
complex-rotation angles $\theta <\pi /4$ is different from the one that
comes from greater angles $\theta >\pi /4$. Epifanov\cite{E96} and Abramov
et al\cite{AAD01} also chose this model for resonance calculations. The
latter authors stated that their results broadly agreed with those of Rittby
et al\cite{REB82}. Andersson\cite{A93} argued that the WKB method with one
transition point is insufficient to calculate the actual resonances beyond
the threshold energy. When adding the necessary transition points their
results broadly agree with those of Rittby et al\cite{REB81,REB82,REB82b}. B%
\"{o}gli et al\cite{BBMTW14} developed a method for enclosing and excluding
resonances with ``guaranteed certainty''. They concluded that some of the
complex eigenvalues obtained by Korsch et al\cite{KLM82} are not true
resonances. For the commonly chosen parameters $J=0.8$, $\lambda =0.1$ the
potential supports only one bound state with energy $E_{0}<J$ and many
resonances.

On studying the performance of the Riccati-Pad\'{e} method (RPM) for the
calculation of bound states and resonances Fern\'{a}ndez\cite{F96} found an
apparently strange resonance located quite close to the only bound state of
the model. This resonance had in fact been reported by Rittby et al\cite
{REB82b} and labelled as the KLM pole $0^{+}$.

The purpose of this paper is to investigate if the\ RPM yields both sets of
poles REB and KLM\cite{REB82b} or just one kind. To this end we carry out
extremely accurate RPM calculations and compare them with the results
provided by the Rayleigh-Ritz method with complex rotation.

\section{The model}

In this paper we study the spectrum of the dimensionless Hamiltonian
operator $H=p^{2}+V(x)$, where $p=-id/dx$ and
\begin{equation}
V(x)=\left( x^{2}-2J\right) e^{-\lambda x^{2}}+2J,\;J,\lambda >0.
\label{eq:V(x)}
\end{equation}
Note that this Hamiltonian, which is the one chosen by Fern\'{a}ndez\cite
{F96}, is exactly twice the one mentioned above\cite
{ALR80,REB81,REB82,KLM82,REB82b,E96,AAD01}. The potential (\ref{eq:V(x)})
exhibits a minimum $V(0)=0$ at origin and two barriers of height
\begin{equation}
V(\pm x_{b})=\frac{e^{-2j\lambda -1}}{\lambda }+2J,\;x_{b}=\sqrt{\frac{%
2J\lambda +1}{\lambda }},  \label{eq:V(xb)}
\end{equation}
located at $x=\pm x_{b}$. In addition to it, $\lim\limits_{|x|\rightarrow
\infty }V(x)=2J$ is the threshold of the continuum spectrum. That is to say:
we expect bound states for $0<E<2J$ and unbound states for $\Re E>2J$. It is
well known that there is always a bound state $\psi _{0}(x)$ with energy $%
E_{0}$ for all values of $J>0$. The Hellmann-Feynman theorem tells us that
the bound states satisfy
\begin{equation}
0<\frac{\partial E}{\partial J}=2\left\langle 1-e^{-\lambda
x^{2}}\right\rangle <2.  \label{eq:Hell-Feyn}
\end{equation}
The energies of the bound states increase with $J$ more slowly than the
threshold $2J$ and as $J$ increases more bound states appear.

The Taylor expansion of $V(x)$ about the origin
\begin{equation}
V(x)=\left( 2J\lambda +1\right) x^{2}-\lambda \left( J\lambda +1\right)
x^{4}+\frac{\lambda ^{2}\left( 2J\lambda +3\right) }{6}x^{6}+\ldots
\label{eq:V(x)_Taylor}
\end{equation}
suggests that if $\lambda \ll 1$ the bound-state eigenvalues are
approximately given by $E_{n}\approx \sqrt{2J\lambda +1}(2n+1)$, $%
n=0,1,\ldots $, provided that $E_{n}\ll 2J$. In other words, the harmonic
approximation is valid in the limit of sufficiently small $\lambda $ and
sufficiently large $J$.

\section{The Riccati-Pad\'{e} method}

\label{sec:RPM}

The dimensionless Schr\"{o}dinger equation for a one-dimensional model reads
\begin{equation}
\psi ^{\prime \prime }(x)+\left[ E-V(x)\right] \psi (x)=0,
\label{eq:Schrödinger}
\end{equation}
where $E$ is the eigenvalue and $\psi (x)$ is the eigenfunction that
satisfies some given boundary conditions. For example, $\lim\limits_{|x|%
\rightarrow \infty }\psi (x)=0$ determines the discrete spectrum and the
resonances are associated to outgoing waves in each channel (for example, $%
\psi (x)\sim Ae^{ikx}$).

In order to apply the RPM we define the regularized logarithmic derivative
of the eigenfunction
\begin{equation}
f(x)=\frac{s}{x}-\frac{\psi ^{\prime }(x)}{\psi (x)},  \label{eq:f(x)}
\end{equation}
that satisfies the Riccati equation
\begin{equation}
f^{\prime }(x)+\frac{2sf(x)}{x}-f(x)^{2}+V(x)-E=0,  \label{eq:HO_Riccati}
\end{equation}
where $s=0$ or $s=1$ for even or odd states, respectively. If $V(x)$ is a
polynomial function of $x$ or it can be expanded in a Taylor series about $%
x=0$ then one can also expand $f(x)$ in a Taylor series about the origin
\begin{equation}
f(x)=x\sum_{j=0}^{\infty }f_{j}(E)x^{2j}.  \label{eq:f(x)_series}
\end{equation}
On arguing as in earlier papers (see, for example \cite{F96} and references
therein) we conclude that we can obtain approximate eigenvalues to the
Schr\"{o}dinger equation from the roots of the Hankel determinant
\begin{equation}
H_{D}^{d}(E)=\left|
\begin{array}{cccc}
f_{d+1} & f_{d+2} & \cdots & f_{d+D} \\
f_{d+2} & f_{d+3} & \cdots & f_{d+D+1} \\
\vdots & \vdots & \ddots & \vdots \\
f_{d+D} & f_{d+D+1} & \cdots & f_{d+2D-1}
\end{array}
\right| =0,  \label{eq:Hankel}
\end{equation}
where $D=2,3,\ldots $ is the dimension of the determinant and $d$ is the
difference between the degrees of the polynomials in the numerator and
denominator of the rational approximation to $f(x)$. In those earlier papers
we have shown that there are sequences of roots $E^{[D,d]}$, $D=2,3,\ldots $
of the determinant $H_{D}^{d}(E)$ that converge towards the bound states and
resonances of the quantum-mechanical problem. We have at our disposal many
sequences, one for each value of $d$, but it is commonly sufficient to
choose $d=0$. For this reason, in this paper we restrict ourselves to the
sequences of roots $E^{[D]}=E^{[D,0]}$ (unless stated otherwise).

The Hankel determinants (\ref{eq:Hankel}) are polynomial functions of $E$
with real coefficients. Therefore, since both $E$ and $E^*$ are roots we
simply show the absolute value of the imaginary part of the complex
eigenvalues calculated by means of the RPM.

It has been shown that the quantization condition (\ref{eq:Hankel}) is
consistent with moving a zero of $\psi (x)$ towards infinity either along
the real axis\cite{AB11,FG13a} or along a ray $xe^{i\beta }$ on the complex
coordinate plane\cite{FG13b}. In order to appreciate the latter statement
clearer consider the canonical transformation
\begin{equation}
UxU^{-1}=\gamma x,\;UpU^{-1}=\gamma ^{-1}p,  \label{eq:scaling}
\end{equation}
that is commonly called scaling or dilatation transformation. If $\gamma $
is real, then $U$ is unitary and $U^{-1}=U^{\dagger }$ (the adjoint of $U$).
The coefficients $\tilde{f}_{j}$ of the Taylor expansion of $\tilde{f}%
(x)=f(\gamma x)$ about $x=0$ are given by $\tilde{f}_{j}=\gamma ^{2j+1}f_{j}$
and the corresponding Hankel determinants are related by $H_{D}^{d}(\tilde{f}%
)=\gamma ^{D(2D+2d+1)}H_{D}^{d}(f)$. It is clear from this expression that
the roots of the Hankel determinant $H_{D}^{d}(f)$ are also those of $%
H_{D}^{d}(\tilde{f})$.

\section{Results and discussion}

We first comment on a particular feature of the RPM that was already
discussed in earlier papers(see, for example, \cite{F96}). The canonical
transformation (\ref{eq:scaling}) with $\gamma =e^{i\theta }$ leads to
\begin{equation}
UHU^{-1}=e^{-2i\theta }\left[ p^{2}+e^{2i\theta }V(e^{i\theta }x)\right] .
\label{eq:UHU+}
\end{equation}
When $\theta =\pi /2$ then
\begin{equation}
UHU^{-1}=-H_{CR},\;H_{CR}=p^{2}+\left( x^{2}+2J\right) e^{\lambda x^{2}}-2J.
\label{eq:UHU+_pi/2}
\end{equation}
The Hamiltonian $H_{CR}$ exhibits discrete spectrum for all $E>0$ and,
according to the discussion of the preceding section, the application of RPM
to $H$ yields also the eigenvalues of $-H_{CR}$. For example, from a
sequence of negative roots $E^{[D]}$, $2\leq D\leq 7$, we obtained $%
-E_{0}^{CR}=-1.144507971437882$. Note that in this case the RPM is moving
the zero of $\psi (x)$ towards infinity along the imaginary axis ($%
UxU^{-1}=ix$).

Some time ago, Rittby et al\cite{REB81,REB82} calculated the resonances for
the potential (\ref{eq:V(x)}) with $J=0.8$ and $\lambda =0.1$ finding a
curious oscillation in the plot of $\Re E$ vs. $\Im E$ and that $\Re
E<E_{threshold}$. Korsch et al\cite{KLM82} argued that such oscillation was
due to numerical instabilities or to a limited range of variation of the
complex-rotation angle and presented alternative results for $\Re E$ vs. $%
\Im E$ that exhibited a smoother behaviour with a maximum. The discrepancy
was found to be more noticeable between the resonances with high quantum
number. In a reply to this comment Rittby et al\cite{REB82b} showed that one
obtains either one set of results or the other depending on the angle of
rotation of the coordinate in the complex plane. They obtained their earlier
results when $\theta <\theta _{crit}$ and those of Korsch et al\cite{KLM82}
when $\theta >\theta _{crit}$, where $\theta _{crit}=\frac{\pi }{4}$ is the
angle at which the asymptotic limit of $V(e^{i\theta }x)$ ceases to exist.
More precisely, the real part of $V(e^{i\theta }x)$ exhibits an oscillation
of increasing magnitude when $\theta \geq \frac{\pi }{4}$.

It follows from the discussion above that there are two sets of eigenvalues
that for brevity we decided to call type $a$ and type $b$. The former appear
at complex-rotation angles $\theta <\frac{\pi }{4}$ and the latter at $%
\theta >\frac{\pi }{4}$. They are obviously the REB and KLM poles discussed
by Rittby et al\cite{REB82b} and reported in their Tables I and II,
respectively. The RPM yields both sets of resonances but those of type $a$,
including the bound state that is probably the REB pole $0^{+}$, appear at
considerably larger determinant dimensions. For example, from determinants
of order $115\leq D\leq 132$ we estimated
\begin{equation}
E_{16}^{a}=9.19265185-24.2859880i,
\end{equation}
while, on the other hand, from determinants of dimension $D\leq 34$ we
obtained
\begin{equation}
E_{16}^{b}=9.178238697954503583761-24.263016247192105546239i.
\end{equation}

For even solutions $\psi (-x)=\psi (x)$ there is always a bound state and
from roots of Hankel determinants of order $D\leq 34$ we obtained
\begin{equation}
E_{0}^{bs}=1.004080724283934.
\end{equation}
As stated above, this bound state is probably the REB pole $0^{+}$ that was
supposed to exhibit a very small imaginary part ($\sim 10^{-14}$)\cite
{REB82b}. It was also reported in a table of another paper by the same
authors\cite{REB82}. Close to this bound state lays the resonance $E_{0}^{b}$
that one easily obtains by means of the RPM. From determinants of dimension $%
D\leq 34$ we obtained
\begin{eqnarray}
E_{0}^{b} &=&1.004080726301570469395614592615994014289250-  \nonumber \\
&&0.2934712718907477714672477215058936\times 10^{-8}i.
\end{eqnarray}
It is worth noting that $\left| \Im E_{0}^{b}\right| $ is of the order of $%
\left| \Re E_{0}^{b}-E_{0}^{bs}\right| $.

The first odd resonance of type $b$ is embedded in the continuum:
\begin{eqnarray}
E_{1}^{b} &=&2.84194189142938641479284813290283093-  \nonumber \\
&&0.11653056177108158006256047430109\times 10^{-3}i.
\end{eqnarray}

By means of the RPM we calculated some of the REB poles (Table~\ref
{tab:J08l01a}) and all the KLM poles (Table~\ref{tab:J08l01b}). Resonances
of type $a$ with larger quantum number $n$ are very difficult to obtain by
means of the RPM because they appear at rather too large determinant
dimensions. However, the results shown in these tables are more accurate
than those reported by Rittby et al\cite{REB81,REB82,REB82b} and Korsch et al%
\cite{KLM82} (note that our results are twice those in references\cite
{REB81,REB82,REB82b,KLM82}).

Resonances in the discrete spectrum also appear for odd solutions provided
that $J$ is large enough. For example, when $J=2$ we have one odd bound
state with energy
\begin{equation}
E_{1}^{bs}=3.203701434562602,
\end{equation}
and its partner resonance
\begin{eqnarray}
E_{1}^{b} &=&3.20370148589618139565563226675496312  \nonumber \\
&&-0.83665793634597482016260533385\times 10^{-8}i,
\end{eqnarray}
both obtained from determinants of dimension $D\leq 34$. In this case we
also appreciate that $\left| \Im E_{1}^{b}\right| $ is of the order of $%
\left| \Re E_{1}^{b}-E_{1}^{bs}\right| $. Note that $\Re E_{1}^{b}$
increased with $J$ but not as fast as $2J$ and, consequently, it crossed the
threshold from the continuum to the discrete spectrum. Our numerical results
suggest that the resonances also satisfy the bound-state condition $%
0<\partial \Re E^{res}/\partial J<2$ and that $\partial \left| \Im
E^{res}\right| /\partial J<0$.

For the same potential parameters we have the ground state
\begin{equation}
E_{0}^{bs}=1.117002075677124853805,
\end{equation}
and its partner resonance
\begin{eqnarray}
E_{0}^{b} &=&1.117002075832116444713357703111286477-  \nonumber \\
&&0.9999285894038481299231357\times 10^{-10}i,
\end{eqnarray}
obtained from determinants of dimension $D\leq 34$.

For small $J$ it is easier to obtain the resonance in the discrete part of
the spectrum than the partner bound state by means of the RPM. This
behaviour tends to be exactly the opposite as $J$ increases.

According to the results of Rittby et al\cite{REB82b} (see also present
tables \ref{tab:J08l01a} and \ref{tab:J08l01b}) the REB and KLM poles with
the same quantum number are almost identical if the resonance number $n$ is
small enough. As $n$ increases the members of each pair move apart. Present
results suggest that if $J$ increases a pair of complex eigenvalues crosses
the threshold $2J$ into the discrete spectrum. The eigenvalue of type $a$
becomes the energy of a bound state ($\Im E^{a}=0$ when $\Re E^{a}<2J$)
while the eigenvalue of type $b$ becomes its accompanying resonance.

\begin{figure}[]
\begin{center}
\includegraphics[width=9cm]{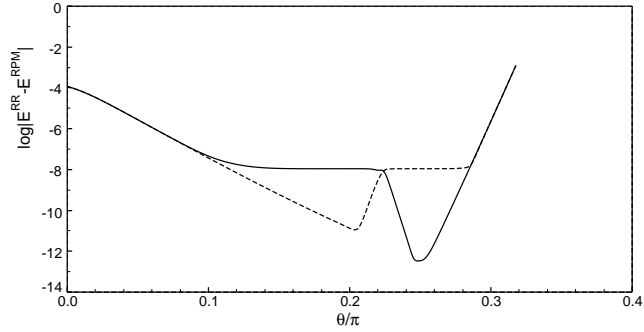}
\end{center}
\caption{$\log\left|E^{RR}(\theta)-E^{RPM}\right|$ for the REB (dash line)
and KLM (solid line) poles when $\lambda=0.1$ and $J=0.8$.}
\label{fig:RR-REB-KLM}
\end{figure}

In order to test the RPM results we have carried out a Rayleigh-Ritz
calculation with complex-rotation (see, for example, reference \cite{MCW78}
and references therein) and the basis set of the harmonic oscillator $%
H_{HO}=p^{2}+x^{2}$. Fig.~\ref{fig:RR-REB-KLM} shows $\log \left|
E^{RR}(\theta )-E_{REB}^{RPM}\right| $ and $\log \left| E^{RR}(\theta
)-E_{KLM}^{RPM}\right| $ for $J=0.8$, $\lambda =0.1$ and $N=80$ basis
functions. This figure shows that the optimal angles satisfy $\theta
_{REB}<\pi /4<\theta _{KLM}$. A more extensive calculation with several
values of $N$ suggests that both optimal complex-rotation angles increase
with $N$ in such a way that while the REB one remains smaller that $\pi/4$
the KLM one becomes clearly greater than such critical angle.

\begin{figure}[]
\begin{center}
\includegraphics[width=6cm]{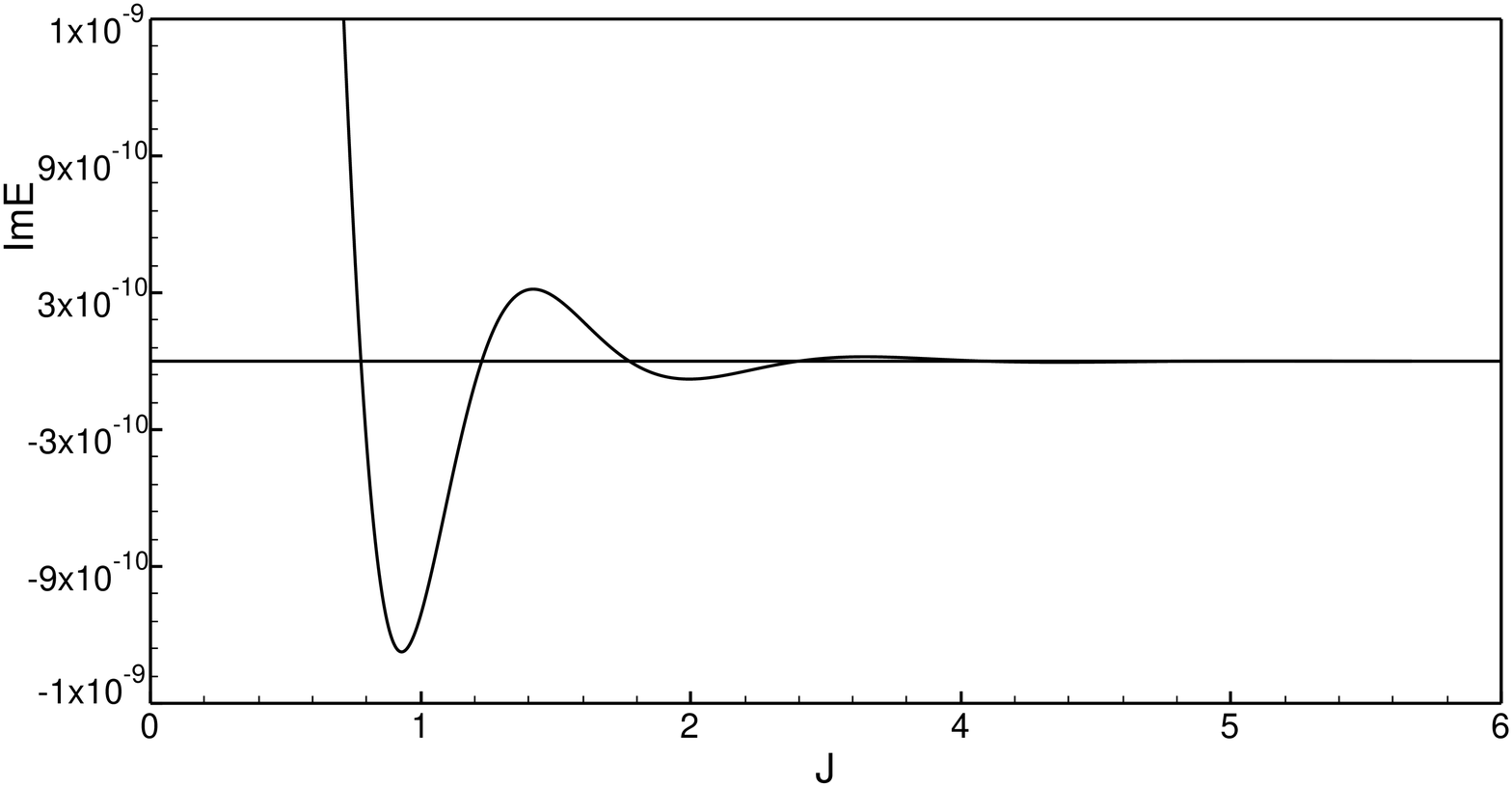} %
\includegraphics[width=6cm]{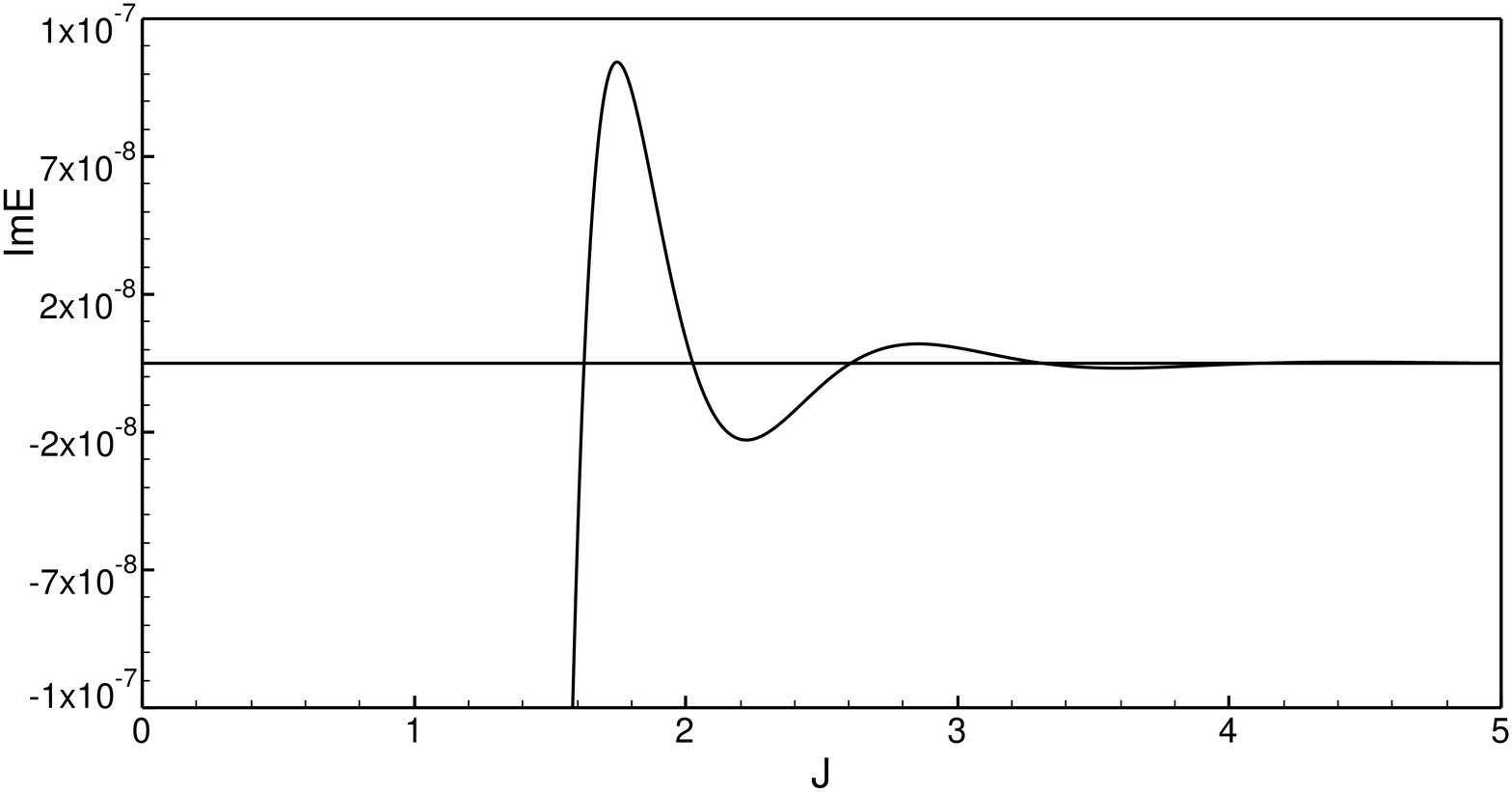}
\end{center}
\caption{$\Im E$ vs. $J$ for the first (left) and second (right) resonances
of type $b$ (KLM poles) }
\label{fig:ImEvJ}
\end{figure}

An interesting property of the resonances of type $b$ (KLM poles) emerged
during the calculation. If we look for stable eigenvalues roughly in the
interval $0.85<\theta <0.95$ then $\Im E^b$ oscillates as shown in Fig.~\ref
{fig:ImEvJ} for the first two ones $E_{0}^{b}$ and $E_{1}^{b}$. On the other
hand, $\Im E^a$ is always negative when $0.65<\theta <0.78$. As argued
above, the latter eigenvalues become real when crossing the continuum
threshold $\Re E=2J$ and the rate of convergence of the Rayleigh-Ritz method
becomes remarkably small about such point.

There is no doubt that the one-dimensional potential (\ref{eq:V(x)})
exhibits two kinds of resonances (REB and KLM poles) that the
complex-rotation method reveals at two different intervals of rotation
angles. What is most interesting is that the RPM yields both sets of
eigenvalues as roots of the same Hankel determinants. The only difference is
that the KLM poles appear in Hankel determinants of smaller dimension and we
can calculate them more accurately when $J$ is relatively small. Exactly the
opposite is commonly true for sufficiently large values of $J$. The RPM
yields both sets of eigenvalues because the roots of the Hankel determinants
are invariant under complex-rotation of the coordinate. Since the resonances
of type $a$ become bound states when they pass from $\Re E^{a}>2J$ to $\Re
E^{a}<2J$ one may interpret them as the usual metastable states and bound
states. It only remains to know if the resonances of type $b$ have any
useful physical meaning. They probably correspond to boundary conditions
different from those of type $a$ but the RPM does not provide such piece of
information.

\begin{table}[]
\caption{Resonances of type $a$ (REB poles) for the potential well (\ref
{eq:V(x)}) with $J=0.8$ and $\lambda=0.1$}
\label{tab:J08l01a}
\begin{center}
\begin{tabular}{D{.}{.}{2}D{.}{.}{20}D{.}{.}{20}}
\hline
 \multicolumn{1}{c}{$n$} & \multicolumn{1}{c}{$\Re E$} &\multicolumn{1}{c}{$|\Im E|$}\\
\hline

0 &  1.00408072428393443017 &                           \\
1 &  2.84194190210246090571 &  0.00011653325419685182  \\
2 &  4.25439414535445676474 &  0.03089463756140796363  \\
3 &  5.16916573799994004827 &  0.34750141927735930069  \\
4 &  5.84884378317999747884 &  1.12958996483545345776  \\
5 &  6.51097253363998538888 &  2.22306318914049287816  \\
6 &  7.11443165024522044127 &  3.51101211133329168976  \\
7 &  7.64865900791597156098 &  4.97489236442085409173  \\
8 &  8.11086942948812965998 &  6.59728208929395179151  \\
9 &  8.49991012723345008717 &  8.36633927847726677570  \\
10&  8.81554505392263084583 &   10.27290632674290915601 \\

\end{tabular}
\end{center}
\end{table}

\begin{table}[]
\caption{Resonances of type $b$ (KLM poles) for the potential well (\ref
{eq:V(x)}) with $J=0.8$ and $\lambda=0.1$}
\label{tab:J08l01b}
\begin{center}
\par
{\scriptsize
\begin{tabular}{D{.}{.}{2}D{.}{.}{20}D{.}{.}{20}}
\hline
 \multicolumn{1}{c}{$n$} & \multicolumn{1}{c}{$\Re E$} &\multicolumn{1}{c}{$|\Im E|$}\\
\hline

0 &  1.00408072630157046940 & 0.00000000293471271891         \\
1 &  2.84194189142938641479 & 0.00011653056177108158        \\
2 &  4.25439415504499186371 & 0.03089462568361036622        \\
3 &  5.16916571970620038273 & 0.34750143832439856191        \\
4 &  5.84884385847547449718 & 1.12958993116515299773        \\
5 &  6.51097228004676307937 & 2.22306320004939896286        \\
6 &  7.11443232530273964386 & 3.51101246935385004749        \\
7 &  7.64865805373778059202 & 4.97489030645579846012        \\
8 &  8.11086733641836896565 & 6.59728840290639287908        \\
9 &  8.49992787752865274035 & 8.36633165517743824541        \\
10&  8.81549677260886623210 & 10.27287881393211008638       \\
11&  9.05762805573781967843 & 12.30961933401526496646       \\
12&  9.22657497347881017987 & 14.47051165461491434216       \\
13&  9.32269370788061645446 & 16.75044181031167253591       \\
14&  9.34639100651463929862 & 19.14500056419140530520       \\
15&  9.29809501050649218041 & 21.65033039282456000024       \\
16&  9.17823869795450358376 & 24.26301624719210554624       \\
17&  8.98725046024366224546 & 26.98000499389483811828       \\
18&  8.72554882720201232788 & 29.79854439102831848701       \\
19&  8.39353964985405639416 & 32.71613581097854432033       \\
20&  7.99161475460693463976 & 35.73049690934641949224       \\
21&  7.52015148083891622536 & 38.83953166339397165944       \\
22&  6.97951274517891567252 & 42.04130598427444775492       \\
23&  6.37004741335819685837 & 45.33402762088381309024       \\
24&  5.69209084394630221200 & 48.71602942021668615706       \\
25&  4.94596551967542290825 & 52.18575524932345994550       \\
26&  4.13198171411806152281 & 55.74174805382654044018       \\
27&  3.25043816167984493764 & 59.38263965038260687933       \\
28&  2.30162271167951448456 & 63.10714194024298827641       \\
29&  1.28581295544208483457 & 66.91403929790148766545       \\
30&  0.20327682052002685664 & 70.80218193928122646417       \\
31&  -0.94572687055748570093& 74.77048011249426351515       \\
32&  -2.16094787755166857383& 78.81789898404686559215       \\
33&  -3.44214405508360832566& 82.94345411668353883653       \\
34&  -4.78908089905121521519& 87.14620745346424210130       \\
35&  -6.20153112503278609622& 91.42526373731795161905       \\
36&  -7.67927427662588964625& 95.77976730707267051567       \\
37&  -9.22209636161676757971&  100.20889922046949752260      \\
38&  -10.82978951390703259870 &   104.71187466241100492439  \\
39&  -12.50215167920702077572 &   109.28794060304091487333  \\
40&  -14.23898632261633438011 &   113.93637367548645768060  \\

\end{tabular}
}
\end{center}
\end{table}

\end{document}